\documentclass[]{JFM-FLM_Au}

\usepackage{booktabs}
\usepackage{multirow}
\usepackage{bigdelim} 
\usepackage{amsmath} 
\usepackage{subcaption}

\lefttitle{Ron Shnapp and Stefano Brizzolara}
\righttitle{Journal of Fluid Mechanics}

\title{Lagrangian Proper Orthogonal Decomposition}

\author{Ron Shnapp\aff{1} and Stefano Brizzolara\aff{2}}

\affiliation{\aff{1}Mechanical Engineering Department, Ben-Gurion University of the Negev, Beer-Sheva, Israel
\aff{2}Department of Mechanical and Aerospace Engineering, Princeton University, Princeton, NJ, USA}

\corresau{Stefano Brizzolara, s.brizzolara@princeton.edu}

\begin{document}
\maketitle

\begin{abstract}
We introduce a modal representation for Lagrangian trajectories in turbulence, termed Lagrangian Proper Orthogonal Decomposition (LPOD). An ensemble of particle trajectories is used to construct velocity time series, which are normalized independently for each trajectory to isolate fluctuations. Principal Component Analysis is then applied to the resulting dataset, with temporal instances defining the feature space. The method is tested on trajectories from both direct numerical simulations of homogeneous isotropic turbulence and three-dimensional particle-tracking experiments, showing that the leading modes exhibit similar structures and energy distributions in both cases. Truncated reconstructions are obtained by combining modes and coefficients, rescaling the fluctuations, and integrating in time. For trajectories of the order of the integral time scale, single-particle dispersion and curvature statistics are accurately reproduced using a limited number of modes ($\approx$ 10), whereas capturing the tails of acceleration distributions requires a larger set ($\approx$ 30–60). Longer trajectories require progressively more modes for accurate reconstruction. These results suggest a possible route to data-driven generation of synthetic particle trajectories via stochastic sampling of the modal Lagrangian dynamics.
\end{abstract}


\section{Introduction}
\label{sec:Introduction}
Proper Orthogonal Decomposition (POD) is a widely used technique for extracting coherent structures from complex flow data. Originally introduced in fluid mechanics by \citet{lumley1967structure}, it decomposes a flow field into a set of orthogonal modes that capture most of the kinetic energy. In this framework, a fluctuating velocity field is expressed as a sum of spatial, orthogonal modes, each modulated by time-dependent coefficients, whose order is determined by their energy content \citep{weiss2019tutorial}. In its discrete form, POD is equivalent to Principal Component Analysis (PCA), which consists in solving an eigenvalue problem for the data covariance matrix~\citep{karhunen1947lineare}.

POD has been primarily developed and applied within an Eulerian framework. In this setting, the method is applied to a set of flow realizations, which correspond to instantaneous snapshots of the velocity field measured at fixed spatial locations. Each realization is represented by the spatial distribution of the velocity field, including all components, arranged into a single state vector. The decomposition is obtained by solving the eigenvalue problem associated with the data covariance operator. The resulting eigenvectors define an orthogonal set of spatial modes, while the corresponding eigenvalues quantify their kinetic energy. Two equivalent formulations of the Eulerian POD are commonly used. In the first -- the snapshot method \citep{sirovich1987turbulence} -- the covariance matrix is constructed from correlations between different realizations. In the second case, the covariance is constructed in physical space, thereby correlating spatial degrees of freedom across the dataset.

Despite POD's widespread use in Eulerian problems in fluid mechanics, applications in Lagrangian settings -- where fluid properties are prescribed along fluid parcel positions \citep{monin2013statistical} -- remain limited. \citet{schiodt2022characterizing} developed a particle-POD (PPOD) method for inertial particles in decaying turbulence. \citet{hassanian2025proper} attempted to compare the distribution of POD mode energy from a Lagrangian experiment to Kolmogorov spectra, although how these are related is not yet clear. In the context of Lagrangian dynamics, \citet{shinde2021lagrangian} developed a POD routine called Lagrangian modal analysis (LMA) that decomposes Lagrangian fields (the flow map) to identify coherent structures using space as a variable. The authors demonstrated that LMA can identify Lagrangian coherent structures in a manner similar to finite-time Lyapunov exponent analysis. In other contexts, PCA has been used to analyze particle trajectories, e.g., in soft matter; for example, \citet{chen2015principal} used PCA to analyze particle trajectories, but at the image level rather than as a single-trajectory modal analysis.

Since the introduction of Lagrangian experimental and numerical techniques roughly 30 years ago \citep{yeung1989lagrangian,maas1993particle,malik1993particle}, the Lagrangian viewpoint has revealed numerous important phenomena related to dispersion \citep{biferale2005lagrangian, tan2022universality, shnapp2018generalization}, energy transfer \citep{meneveau1994lagrangian, pumir2001lagrangian} and intermittency \citep{biferale2008lagrangian, benzi2010inertial} in turbulence, as well as properties of inertial particles therein \citep{bec2010turbulent}. Nevertheless, our understanding of how the nonlinear multiscale interactions inherent to turbulent flows lead to these phenomena remains in its infancy. This work is motivated by the notion that a Lagrangian modal decomposition could help in this respect.  

In this paper, we introduce a method, \textit{LPOD}, to decompose Lagrangian trajectories from a turbulent flow into orthogonal modes. We demonstrate its applicability using trajectories from both a direct numerical simulation (DNS) of homogeneous and isotropic turbulence (HIT) and a three-dimensional particle tracking (3D-PTV) experiment of quasi-HIT in a von-Karman flow. With an appropriate data arrangement and normalization, relatively long trajectories can be reconstructed using a limited number of modes, which appear to have a similar functional form in both the DNS and the experimental data. We also show that the method faithfully captures the intermittency of Lagrangian acceleration, velocity differences, and curvature. The method has the potential to advance modeling and data compression in the context of the Lagrangian description of turbulent flows.

\section{\label{sec:Methods} Methods}
\subsection{\label{subsec:The LPOD routine} The LPOD routine}
We consider the Lagrangian trajectories of tracer particles of a homogeneous, isotropic, and stationary turbulent flow, which are defined as the solutions of the following equation:
\begin{equation}
\frac{d\textbf{x}}{dt}=\textbf{v}(\textbf{x}(t),t), \qquad \textbf{x}(t_0)=\textbf{x}_0,
\end{equation}
where $\textbf{x}_0$ and $t_0$ denote the initial position and initial time, respectively, and $\textbf{v}$ is the Eulerian velocity field. The corresponding Lagrangian velocity is given by $\textbf{u}(\textbf{x}_0,t_0;\tau) = \textbf{v}(\textbf{x}(t_0+\tau),t_0+\tau)$, where $\tau=t-t_0$ is the time lag along the trajectory. The assumptions of statistical homogeneity and stationarity imply that the statistics of the Lagrangian velocity do not depend on $\textbf{x}_0$ or $t_0$, so these dependencies are henceforth omitted. A discrete ensemble of $N_p$ trajectories is considered, each represented by the velocity signal:
\begin{equation} \label{eq:velocityPosition}
    \frac{d\textbf{x}^{(p)}}{dt} = \textbf{u}^{(p)}(\tau)\textrm{,} \qquad p=1,\dots,N_p \textrm{,} \qquad \tau\in[0,T]
\end{equation}
where $p$ labels the trajectories in the ensemble and $T$ is the duration of all trajectories. For each trajectory $p$ and velocity component $i=1,2,3$, we define the temporal mean and standard deviation:
\begin{align}
\mu_i^{(p)}=\frac{1}{T}\int_0^T u_i^{(p)}(\tau)\,d\tau = \frac{x_i^{(p)}(T) - x_i^{(p)}(0)}{T}\textrm{,} \\
\sigma_i^{(p)} =
\left( \frac{1}{T}\int_0^T \left(u_i^{(p)}(\tau)-\mu_i^{(p)}\right)^2 d\tau
\right)^{1/2}\textrm{.}
\end{align}
The normalized velocity fluctuation is then defined as:
\begin{equation}
\tilde{u}_i^{(p)}(\tau) =
\left(u_i^{(p)}(\tau)-\mu_i^{(p)}\right)/\sigma_i^{(p)}\textrm{.}
\end{equation}
Notably, as $\mu_i^{(p)}$ is defined per particle, the normalized velocity fluctuations do not impose any net displacement across the time $T$.

The LPOD is defined as a Karhunen--Lo\`eve expansion of these normalized Lagrangian velocity fluctuations,
\begin{equation}
\tilde{u}_i^{(p)}(\tau) = \sum_{k=1}^{\infty} a_{i,k}^{(p)}\,\phi_k(\tau),
\end{equation}
where $\phi_k(\tau)$ are temporal modes and $a_{i,k}^{(p)}$ are the corresponding modal coefficients. The modes satisfy the orthogonality condition, namely 
$\int_0^T \phi_k(\tau)\,\phi_{\ell}(\tau)\,d\tau = \delta_{k\ell}$. A truncated reconstruction using the first $K$ modes is obtained as
\begin{equation}
u_{i}^{(p,K)}(\tau) = \mu_i^{(p)} + \sigma_i^{(p)} \sum_{k=1}^{K} a_{i,k}^{(p)}\,\phi_k(\tau)\textrm{,}
\end{equation}
and the corresponding trajectory obtained by the truncated reconstruction is then computed by time integration:
\begin{equation}
\textbf{x}^{(p,K)}(\tau) = \textbf{x}^{(p)}(0) + \int_0^\tau \textbf{u}^{(p,K)}(s)\,ds.
\end{equation}

In practice, the velocity is available at discrete times $\tau_n=n\Delta \tau$, with $n=1,\dots, N_t$. The expansion is therefore approximated by applying PCA to the matrix of sampled normalized signals $U_{(p,i),n}=\tilde{u}_i^{(p)}(\tau_n)$ whose rows correspond to staked trajectory velocity components and whose columns correspond to time samples. The eigenvectors of the temporal covariance matrix provide the discrete approximation of the modes $\phi_k(\tau_n)$, while the projections of the signals onto these modes give the coefficients $a_{i,k}^{(p)}$.

\subsection{\label{subsec:Datasets description}Datasets description}
We employ data from both a DNS of HIT and a 3D-PTV experiment. The DNS trajectories are integrated in a $1024^3$ HIT flow, $Re_{\lambda} = 433$, from the Johns Hopkins Turbulence Database -- a widely used open database for exploring turbulence dynamics \citep{li2008public}. Trajectories are $688$ frames long and are integrated with a time step of $\Delta \tau = 0.045 \tau_{\eta}$, where $\tau_{\eta}$ is the Kolmogorov time-scale. The total length of the trajectories corresponds to $\tau_{max} = 30.77 \tau_{\eta} = 0.68 T_I$, where $T_I$ is the integral time scale of the flow. Trajectory data were downloaded directly from the online database via a MATLAB interface. Detailed information about this open database is available from \citet{li2008public}.

The experimental trajectories are obtained from a low-seeding 3D-PTV experiment performed in a von Karman turbulence tank. The trajectories have been reconstructed using the MyPTV open-source software \citep{shnapp2022myptv}. Trajectories are $200$ frames long and are sampled with a time step of $\Delta \tau = 0.014 \tau_{\eta}$. The total length of the trajectories corresponds to $\tau_{max} = 28.65 \tau_{\eta} = 1.04 T_I$. Further details about the experimental dataset are available in the supplementary material.

\subsection{\label{subsec:Statistical observables} Statistical observables}
We introduce a set of statistical observables to assess the performance of truncated LPOD trajectory reconstructions, including modal properties, single-particle Lagrangian statistics, and geometric characteristics of the trajectories.

For the modal representation, we compute the projection coefficients associated with each mode, along with the corresponding variance explained, which quantifies the energy content of each principal component. To characterize Lagrangian transport, we first consider two dispersion-related observables: the mean-squared particle displacement and the deviation from ballistic motion, eq.~\eqref{eq:dx2} and \eqref{eq:dbx2}, respectively:
\begin{subequations}\label{eq:dispersion}
\begin{gather}
\langle \delta \textbf{x}^2 \rangle = \langle \left[\textbf{x}(\tau) - \textbf{x}_0\right]^2 \rangle \label{eq:dx2} \\
\langle \delta_b \textbf{x}^2 \rangle = \langle [\textbf{x}(\tau) - (\textbf{x}_0 + \textbf{u}(0)\,\tau) ]^2 \rangle \label{eq:dbx2}
\end{gather}
\end{subequations}
The mean-squared displacement exhibits ballistic scaling for $\tau \ll T_I$, namely $\langle \delta \textbf{x}^2 \rangle \sim u_{rms}^2 \tau^2$ \citep{taylor1922diffusion}. The deviation from ballistic motion isolates the contribution of velocity fluctuations \citep{shnapp2019lagrangian} and is expected to scale as $\langle \delta_b \textbf{x}^2 \rangle \sim \epsilon \tau^3$ in the inertial range (see the supplementary material for the derivation of the scaling).

We further consider the second-order Lagrangian longitudinal structure function:
\begin{equation}
    S_2^{\parallel}(\tau) = \langle {\delta u_{\parallel} (\tau) }^2\rangle\textrm{,}
\end{equation}
where the longitudinal velocity increment is defined as the projection of the velocity difference along the displacement direction,
\begin{equation}
    \delta u_{\parallel}(\tau) = (\textbf{u}(\tau) - \textbf{u}(0)) \,\frac{\textbf{x}(\tau) - \textbf{x}_0}{|\textbf{x}(\tau) - \textbf{x}_0|}\textrm{.}
\end{equation}
In the inertial range, this quantity is expected to scale linearly with time, namely $S_2^{\parallel} \sim \epsilon \tau$ \citep{monin2013statistical}

Furthermore, we consider the magnitude of the Lagrangian acceleration, which is defined as the material derivative of the velocity:
\begin{equation}
    a(\tau) = |\textbf{a}(\tau)| \textrm{,} \quad \textbf{a}(\tau) = \frac{d \textbf{u}(\tau)}{d \tau}\textrm{.}
\end{equation}
At sufficiently high Reynolds number, the probability distribution of the Lagrangian acceleration, if normalized by the Kolmogorov scales, is expected to be universal and follow a multifractal distribution \citep{toschi2009lagrangian}.

Finally, we characterize the trajectory geometry through the curvature:
\begin{equation}
    \kappa = \frac{|\textbf{u} \times \textbf{a}|}{|\textbf{v}|^3}\textrm{.}
\end{equation}
At sufficiently high Reynolds number, the curvature distribution exhibits a power-law behavior that scales linearly for small curvatures and as $\kappa^{-2/5}$ for large curvatures \citep{xu2007curvature}.

\section{Results}

\subsection{LPOD modes and their energy}

\begin{figure}[htbp]
\centering
\begin{minipage}{0.45\textwidth}
    \centering
    \begin{subfigure}{\linewidth}
        \includegraphics[width=\linewidth]{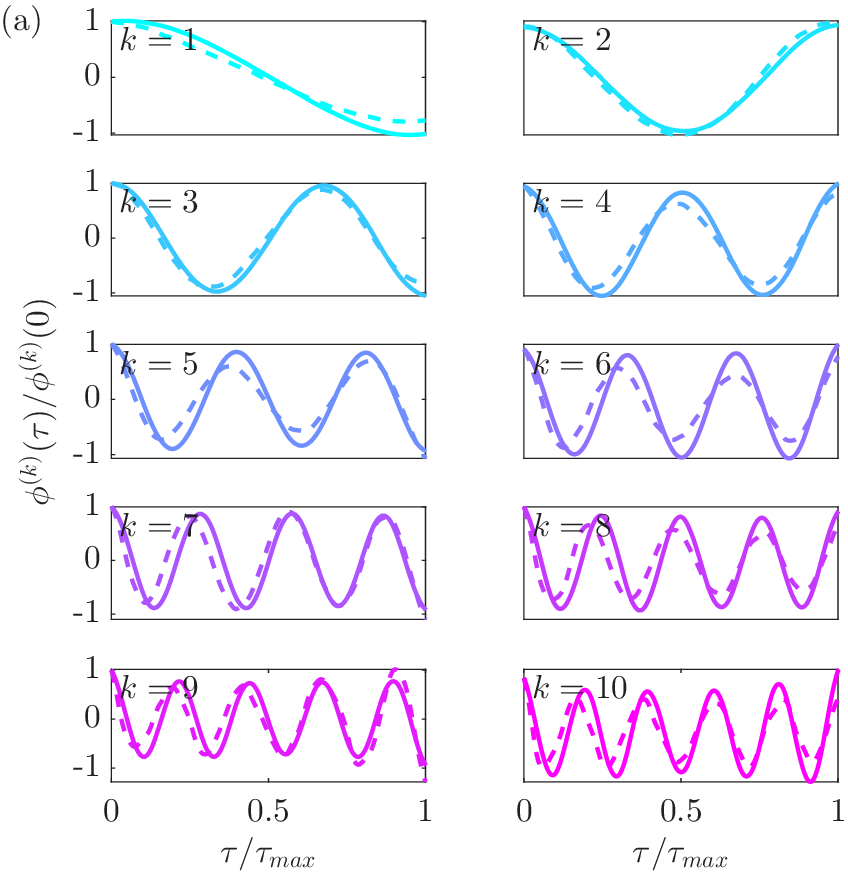}
    \end{subfigure}
    \begin{subfigure}{\linewidth}
        \includegraphics[width=\linewidth]{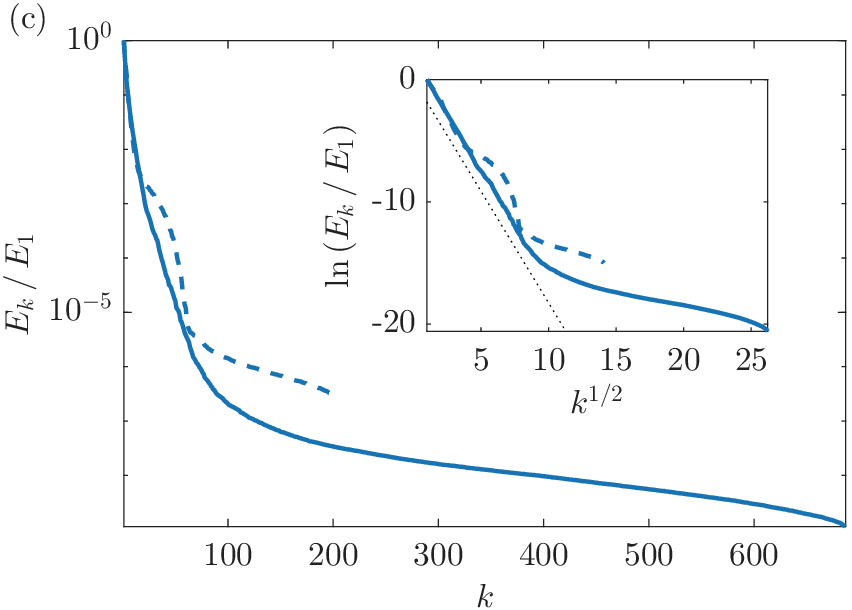}
    \end{subfigure}
\end{minipage}
\hfill
\begin{minipage}{0.5\textwidth}
    \centering
    \begin{subfigure}{\linewidth}
        \includegraphics[width=\linewidth]{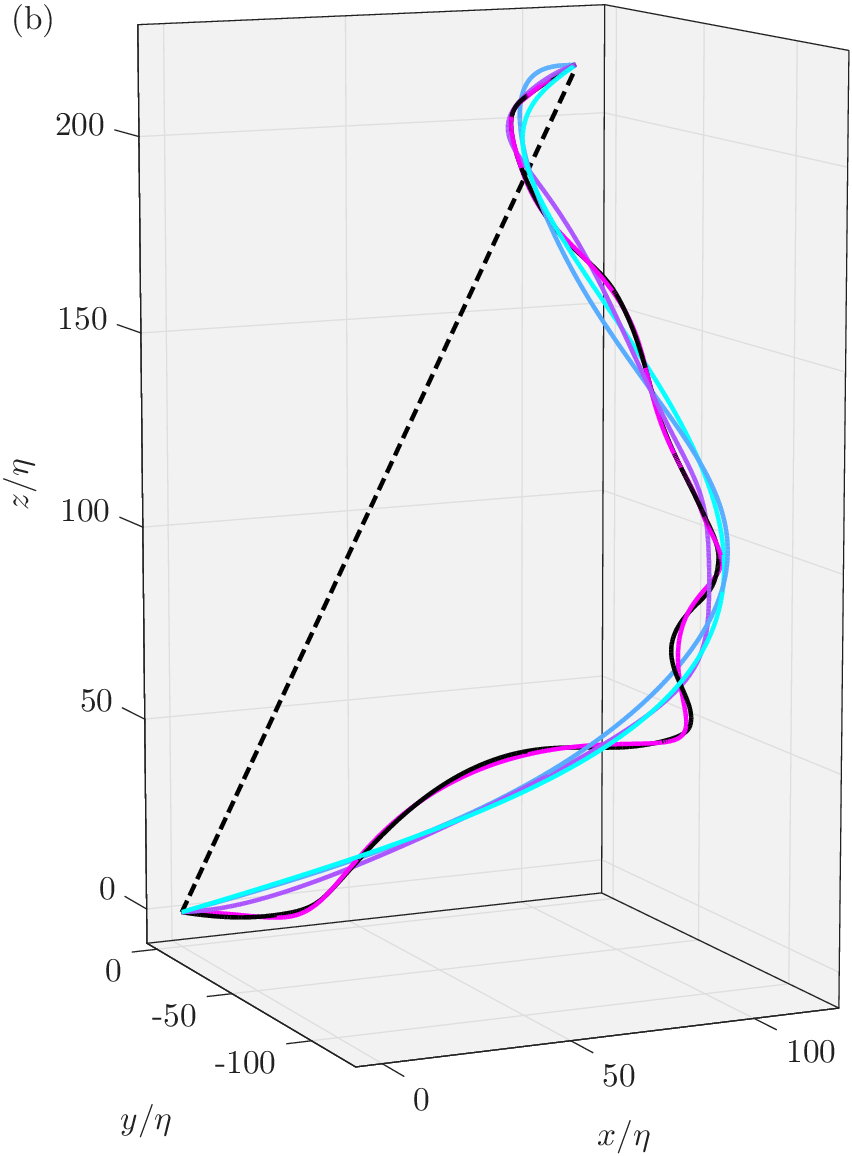}
    \end{subfigure}
\end{minipage}

\caption{(a) First 10 LPOD modes: solid lines denote DNS data and dashed lines 3D-PTV data. (b) Example DNS trajectory reconstruction using an increasing number of modes (2,  4, 8, and 16); the dashed line shows the ballistic reconstruction from the mean velocity, and the black line the original trajectory. (c) Energy distribution normalized such that $\sum_{k=1}^{N_t} K_k = 100$; solid and dashed lines correspond to DNS and 3D-PTV data, respectively. The inset shows the same data in square-log coordinates to highlight the sub-exponential decay.}
\label{fig:modes}

\end{figure}

We start by presenting the results of LPOD applied to both the numerical and experimental datasets described in section \ref{sec:Methods}. Figure \ref{fig:modes} (a) shows the first 10 modes of the expansion, where the time is normalized with the trajectory duration and the intensity with the first value (the value at $\tau = 0$). In both the experiment and the simulation, progressively higher-order modes capture higher-frequency velocity fluctuations. By comparing the leading modes obtained from the experiments (dashed lines) and the simulation (solid line), we observe that the mode shapes are similar in both cases. Figure \ref{fig:modes} (b) shows an example of truncated reconstruction for a topologically complex trajectory from the DNS data. Although truncated at the 16th mode, the reconstruction qualitatively matches the ground truth. The similarity between the modes from the two datasets is an important feature of LPOD, as it has significant potential for data reconstruction, compression, and modeling. This result is, in fact, expected, as a well-known feature of the orthogonal decomposition is that statistically stationary signals decompose into the harmonic basis of a discrete Fourier transform~\citep{George1988}. Since the flow considered here is stationary, homogeneous, and isotropic (nearly so in the experiment), the trajectories are expected to follow stationary statistics, which is encapsulated in the modes we obtained. Deviations from ideal HIT behavior are evident in the differences between the two datasets, which become progressively more pronounced as $k$ increases. These discrepancies arise because the experimental flow is sampled with finite-size particles and the DNS dataset is not perfectly stationary, situations that cannot be represented by purely periodic modes~\citep{hodvzic2024}. Notably, orthogonal decomposition strategies are also very relevant for inhomogeneous or non-stationary flows or those that are homogeneous in some ways but not others, as discussed by~\citet{George1988}. 

Figure~\ref{fig:modes} (c) shows the percentage of energy content associated with each mode (its eigenvalue) for the experiments and DNS data. As with PCA analyses, most of the energy in LPOD is concentrated in the first few modes, with the first 10 modes accounting for 99\% of the kinetic energy. As highlighted by the inset of figure \ref{fig:modes} (c), the mode energy decreases sub-exponentially, approximately as $e^{-(k/C)^{1/2}}$, where $C$ is a constant, up to $k\approx 100$. For $k$ larger than this, the deviation between the experimental and numerical cases becomes more significant. Several factors may lead to discrepancies between datasets. First, Reynolds-number differences and finite experimental resolution lead to differences in the separation of scales, so different behavior is expected when plotting mode energy using a single scaling for all modes. Second, the lengths of the trajectory numbers differ, and, because the number of modes equals the number of time samples per trajectory under the construction method used here, the decompositions for the two datasets are truncated at different values.

\subsection{Reconstructing Lagrangian statistics}
\begin{figure}
\centering
\includegraphics[width=.47\textwidth]{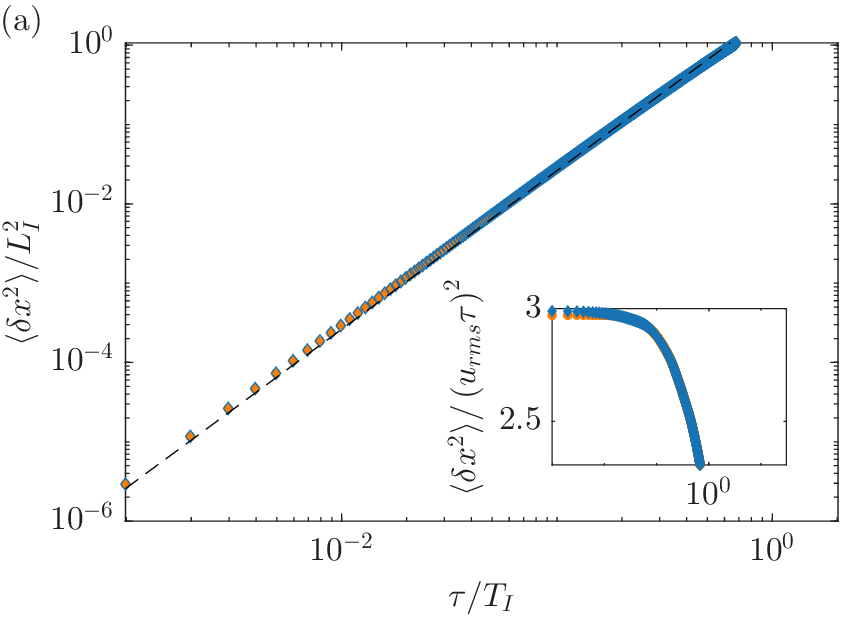}\qquad
\includegraphics[width=.47\textwidth]{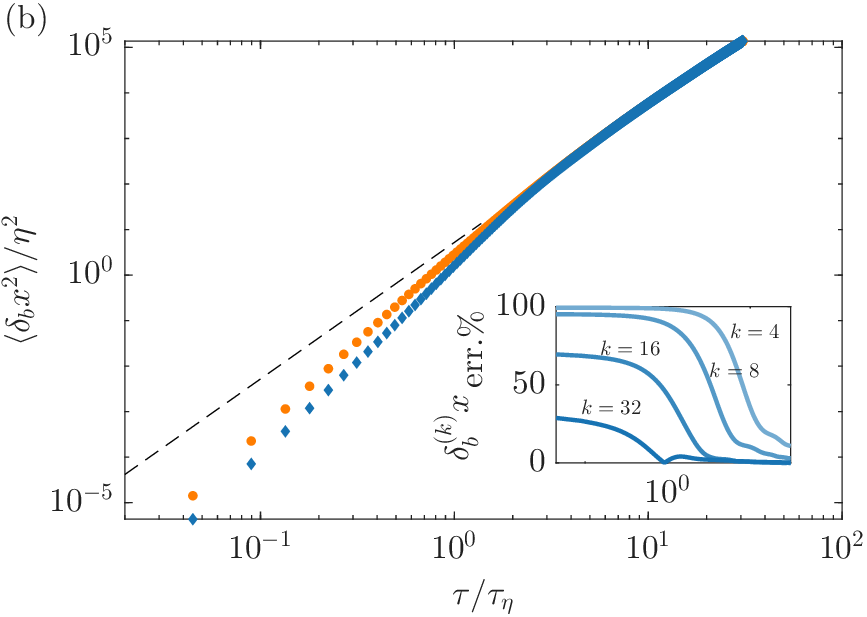}
\caption{(a) Single particle dispersion; the inset shows the data normalized by the ballistic scaling. (b) Deviation from ballistic dispersion; the inset shows the relative reconstruction error $\delta_b^{(k)}x_{\textrm{\,err.\%}} = 100\cdot |\langle \delta_b^{(k)} x ^2 \rangle - \langle \delta_b x ^2 \rangle |  / \langle \delta_b x ^2 \rangle$. Orange denotes the original data, whereas blue denotes the 16 modes truncation.}
\label{fig:dispersion}
\end{figure}

We present the Lagrangian statistics defined in §\ref{subsec:Statistical observables} for both the original and reconstructed trajectories. Figure \ref{fig:dispersion} shows Lagrangian single particle dispersion statistics. Considering the mean-squared displacement, Figure~\ref{fig:dispersion} (a) shows that the truncated reconstruction is indistinguishable from the original signal. This is expected, since the mean velocity is removed prior to the decomposition and reintroduced during reconstruction. As a result, the ballistic contribution to the displacement is preserved exactly, so that, in the ballistic regime, the mean-squared displacement is recovered even in the absence of fluctuating modes. To better assess the limits of the truncated representation on the single-particle dispersion process, we consider the deviation from ballistic motion $\langle\delta_bx^2\rangle$, shown in Figure~\ref{fig:dispersion}(b). This quantity is more sensitive to small-scale velocity fluctuations and is therefore more difficult to reconstruct, particularly at short time lags. Nevertheless, a relatively small number of modes (16) is sufficient to recover the correct behavior for lags exceeding a few Kolmogorov time scales, when the data are expected to exhibit inertial-range scaling. To better assess the impact of the number of modes on the reconstruction of $\langle\delta_bx^2\rangle$, we evaluate the relative reconstruction error, defined as the ratio between the difference of the reconstructed and original signals and the original signal itself. This is shown in the inset of Figure~\ref{fig:dispersion}(b) as a function of both the time lag and the number of retained modes, demonstrating that adding modes reduces the relative error at progressively shorter time lags (smaller scales).

\begin{figure}
\centering
\includegraphics[width=.47\textwidth]{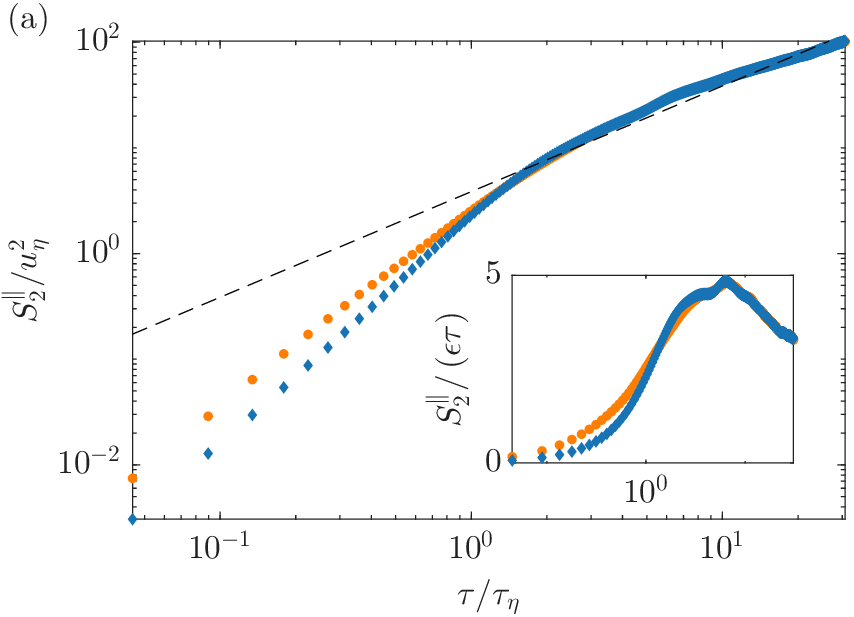}\qquad
\includegraphics[width=.47\textwidth]{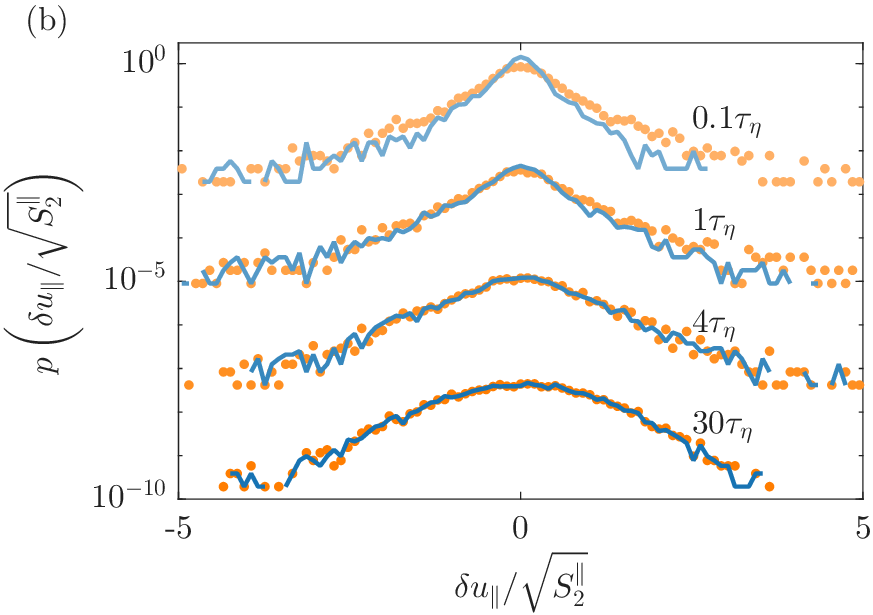}
\caption{(a) Second-order Lagrangian longitudinal structure function computed from the DNS dataset. The inset shows the normalized scaling. Orange circles denote the original data, whereas blue diamonds are obtained from the first 16 modes. (b) Probability density function of the velocity increments from the DNS dataset. The time lag varies from $0.1$ to $30\,\tau_{\eta}$. The reconstructed data (blue solid lines) are obtained using the first 16 modes; the orange circles are the original data. The data are normalized with the structure function computed from the original data for both cases. }
\label{fig:structureFunctions}
\end{figure}

Similarly, the second-order longitudinal structure function is accurately reconstructed for lags larger than a few Kolmogorov time scales (Figure~\ref{fig:structureFunctions}(a)). The inset shows the Lagrangian Kolmogorov constant, obtained by normalizing the structure function by the inertial-range scaling $\varepsilon \tau$. Truncating the modal expansion leads to an extended plateau. Figure \ref{fig:structureFunctions} (b) shows the probability density functions of the normalized velocity increments for increasing time lags. As the lag increases, the distributions approach Gaussian behavior. Reconstructions based on a limited number of modes (16) underestimate the tails at short time lags when the distribution is strongly non-Gaussian, whereas they are accurate for longer time lags as Gaussianity is approached.

\begin{figure}
\centering
\includegraphics[width=.47\textwidth]{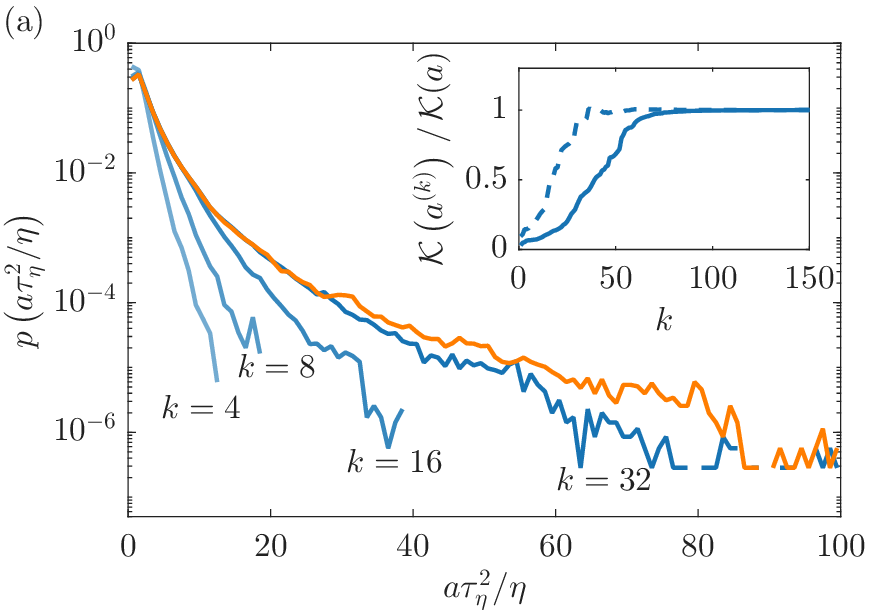}\qquad
\includegraphics[width=.47\textwidth]{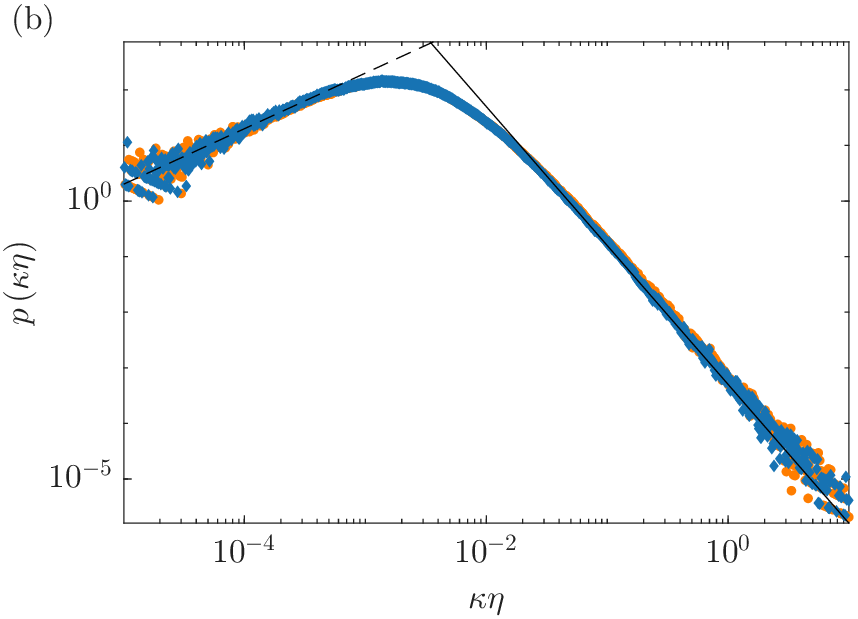}
\caption{(a) Probability density function of the trajectory acceleration. The orange line denotes the original signal, while the blue lines correspond to reconstructions using 4, 8, 16, and 32 modes. The inset shows the kurtosis $\mathcal{K}$ of the reconstructed signal normalized with the original signal Kurtosis, with dashed and solid lines indicating experimental and DNS data, respectively. (b) Probability density function of the trajectory curvature (DNS). The dashed and solid lines indicate the dimensional scalings $\kappa^{1}$ and $\kappa^{-5/2}$ for small and large curvatures, respectively. Data are logarithmically binned; reconstructions use the first 16 modes.}
\label{fig:acc}
\end{figure}

Figure \ref{fig:acc} (a) shows the probability distribution of the trajectory acceleration for the DNS data. The acceleration distribution, as the longitudinal structure function at small time lags, requires high-order modes to be captured accurately, especially in the tail, which accounts for the extreme events. To better highlight this aspect, we compute the ratio of the Kurtosis of the reconstructed acceleration to that of the original data (inset of Figure~\ref{fig:acc}(a)) for both the experimental and numerical data. Up to approximately 60 modes are required to capture the extreme events of the DNS data (continuous line), whereas a smaller number (approximately 30) suffices for the experimental data (dashed line). As such, approximately 0.1\% and 1.0\% of the trajectories' turbulent kinetic energy are needed to resolve acceleration intermittency in the numerical and experimental data, respectively. The difference is likely due to Reynolds number differences and to the trajectories being longer, in terms of Kolmogorov times, for the DNS data than for the experimental data. 

Figure \ref{fig:acc} (b) shows the probability distribution of the trajectory curvature for the DNS data. From this perspective, the reconstructed and original signals appear indistinguishable, indicating that the first 16 modes accurately capture the geometry of the trajectories.

\section{\label{sec:Discussion} Discussion and outlook}
In this paper, we introduced a new method for analyzing Lagrangian trajectories using PCA, which can be viewed as a Lagrangian analog of standard POD. To the best of our knowledge, the only attempt to perform POD on Lagrangian trajectories was by \citet{schiodt2022characterizing}, but with a different approach. In their work, the authors normalized the data with ensemble averages, which are not suited for isolating the fluctuating velocities of each trajectory and do not preserve the total displacement over the full trajectory length. In contrast, our normalization ensures that the truncated reconstructions always begin and end at the same locations as the original trajectories. This is well illustrated by Figure~\ref{fig:modes}(b), which shows that the total drift along each trajectory is preserved by construction, regardless of the number of modes used. Encoding this normalization in the data also renders the single particle dispersion accurately reconstructible with an extremely low number of modes, as shown in Figure~\ref{fig:dispersion}(a), because, for lags shorter than the integral time scale, the process is ballistic \citep{taylor1922diffusion}. However, while this approach preserves the initial and final positions by construction, it does so at the cost of storing the temporal mean and standard deviation for each trajectory separately.

By analyzing the statistics of truncated reconstructions, we show that LPOD's performance depends on the time lag considered. Quantities defined at intermediate and long lags are accurately reconstructed with a small number of modes, whereas observables based on short-time lags, such as acceleration and short-time velocity increments, require a significantly larger number of modes to converge. This reflects the strong intermittency \citep[see, e.g., ][]{mordant2001measurement,la2001fluid} associated with these quantities, which requires higher frequencies and thus higher-order modes to be accurately captured.

Finally, LPOD may serve as a starting point for modeling, particularly for synthetic trajectory generation. This task has been traditionally addressed with stochastic models \citep[][see, e.g., ]{thomson1987criteria, pope1990velocity, sawford1991reynolds, wilson2009lagrangian, shnapp2020} and recently using deep-learning techniques, e.g., diffusion networks \citep[][see, e.g., ]{li2024synthetic,li2025stochastic,li2024generative} or data-informed stochastic models based on neural networks \citep{de2026data}. Applying LPOD to isotropic turbulence, we showed that the shape of the modes appears to only weakly depend on the turbulence under consideration, and may thus be universal -- albeit only two cases (one DNS and one experiment) are reported here; this suggests that learning how to generate the modes' coefficients may be sufficient to construct synthetic Lagrangian turbulence. However, further investigation and testing of non-homogeneous, non-stationary, and non-isotropic turbulent flows are needed to assess the mode shapes in different contexts.

\bibliographystyle{jfm}
\bibliography{jfm}

\end{document}